\documentstyle[sprocl]{article}

\input{psfig}
\def\be{\begin{equation}}
\def\ee{\end{equation}}
\def\bea{\begin{eqnarray}}
\def\eea{\end{eqnarray}}

\begin{document}

\noindent
\begin{minipage}[t]{.2\linewidth}
\leavevmode
 \hspace*{-.8cm}
\psfig{file=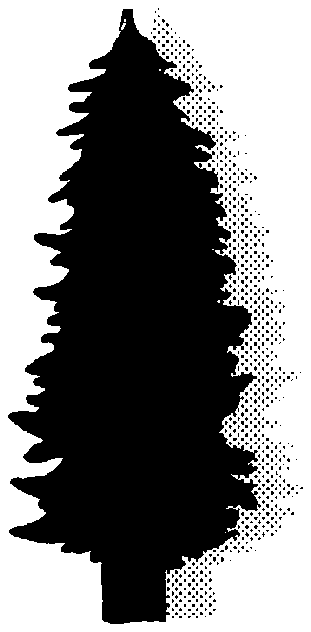,width=3cm}
\end{minipage} \hfill
\begin{minipage}[b]{.45\linewidth}
\rightline{SCIPP 99/38}
\rightline{September 1999}
\vspace{3cm}
\end{minipage}
\vskip1.5cm

\title{TRACKING AND VERTEXING AT A HIGH ENERGY LINEAR COLLIDER}

\author{ B.A. SCHUMM  }

\address{Department of Physics and Santa Cruz Institute of Particle Physics \\
University of California, Santa Cruz, CA 95064, USA }


\maketitle\abstracts{
The relative merits and disadvantages of various alternatives for vertexing
and central tracking detectors for the Linear Collider detector
are presented. Research and development prospects for the various
alternatives are also discussed, as well as a preliminary study of the
prospects for forward ($|\cos \theta| > 0.9$) tracking.
}

\vfill
\begin{center}
{\small Talk presented at ``LCW99", the International Workshop on
Linear Colliders \\
Sitges, Barcelona, Spain, April 28--May 5, 1999}
\end{center}

\vfill\eject
  
\section{Introduction}
Over the last few decades, the design and construction of tracking systems
for cylindrical geometry $e^+e^-$ colliding beam detectors
has become a very advanced and refined field. Nevertheless,
the proposal to build a High Energy Linear Collider (LC), with
a center-of-mass energy as high as $1-1.5$ TeV, and the physics that
would motivate such a facility, combine to present a number of
challenges and opportunities for the design of the associated
charged particle tracking system. This report will summarize
and provide perspectives on recent R\&D accomplishments
and suggestions for
future R\&D, presented at the 1999 International Workshop on
Linear Colliders, which are geared towards meeting these challenges
and exploiting the unique opportunities provided by the Linear Collider
environment.

\section{Motivation for Precise Track Reconstruction}

The physics which motivates the Linear Collider demands
an excellent tracking system. The desire to perform reliable final-state
flavor tagging places a strong demand on the precise measurement
of the track origin parameters. For example, the
$H \rightarrow c {\overline c}$
branching fraction, which is about $6 \times 10^{-2}$ for the Standard
Model Higgs boson for $m_H < 125$ GeV/c$^2$, and less for larger $m_H$,
is an important signal for distinguishing between the Standard Model
Higgs and that of extensions or alternatives to the Standard Model.
While inclusive charm tagging has been done at the $Z^0$ pole
at LEP and the SLC, current tracking systems
would not be able to fully exploit the information available at
a Linear Collider, and thus improvement is desirable.
As another example, in order to fully exploit the natural
$Z^0$ width in the measurement of Higgs properties in the reaction
$e^+e^- \rightarrow Z^0 H$; $Z^0 \rightarrow \mu^+ \mu^-$, an
asymptotic ($p \rightarrow \infty$) momentum resolution
of $\sigma_p/p^2 \sim 5 \times 10^{-5}$
is required.

\section{Linear Collider Backgrounds}

A substantial amount of thinking about LC tracking systems
has been driven by potential machine backgrounds. In the intense
field associated with the $e^+e^-$ collision point, beamstrahlung
photons convert to form $e^+e^-$ pairs. Since the $p_{\perp}$
scale of this process is that of the electron mass, these pairs
can be curled up with a strong solenoidal magnetic field --
a design criterion that is compatible with precise momentum
reconstruction. Although studies continue, it is now thought
that a field of 3-4 T is sufficient to admit the use of a
1 cm beampipe, which is advantageous for precise vertexing.

Somewhat more difficult to model, the tracking system is expected
to be bathed in photons from unconverted beamstrahlung radiation
that scatters off of the apertures of the machine lattice and
masking elements. This background component did much to limit
both the instantaneous and integrated luminosity of the SLC
by producing uncomfortably high occupancy
in the SLD drift chamber as the beam divergence angle was
increased to reduce the spot size, or whenever the orbit
of the beams in the machine went slightly out of tolerance.
A tracking system with the highest granularity in both $r-\phi$
and $r-z$ will be most immune to this background.

As for its hadronic counterparts, minimum bias (`minijet') events
begin to become an issue at the LC. Current estimates~\cite{daSilva}
are that a 500 GeV LC event will have a few-percent probability of
being coincident with a minijet event.
For both the TESLA design, with
$\sim 3,000$ bunch crossings per train separated by 337 nsec each,
and the JLC/NLC design, with $\sim 100$ crossings per train separated
by 2.8 nsec, it would be desirable to have timing which would permit
the rejection of tracks (or hits from the photon background) which
emanate from the wrong bunch. Thus, `temporal segmentation' is
also important, and particularly challenging in the case of the
JLC/NLC.

Finally, neutron halo from the beam dump and collimators, with an
estimated fluence of $2 \times 10^{9}/{\rm cm}^2/{\rm yr}$ for the
inner detectors in the JLC/NLC design, raises some concerns about
radiation damage.

\begin{figure}
\psfig{figure=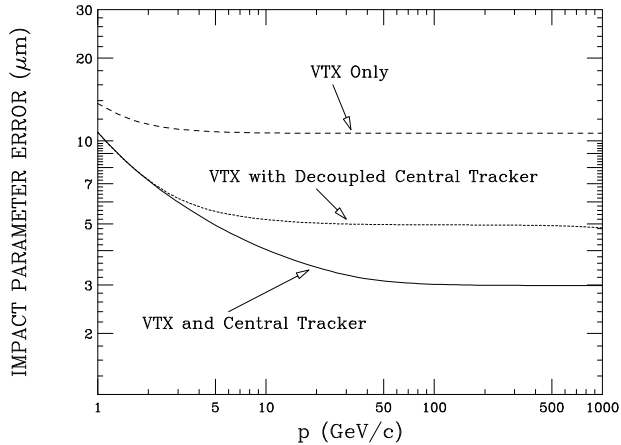,height=2.3in}
\caption{Contributions of the central and inner (VTX) trackers to the
$r-\phi$ impact parameter resolution 
for the `S' detector.
For the case of the
`decoupled' central tracker, an infinite multiple scattering
layer has been inserted between the inner and central trackers,
so that for the central tracker
only the curvature measurement contributes to the
track parameter determination.
\label{fig:cti}}
\end{figure}

\begin{figure}[h]
\centering
\psfig{figure=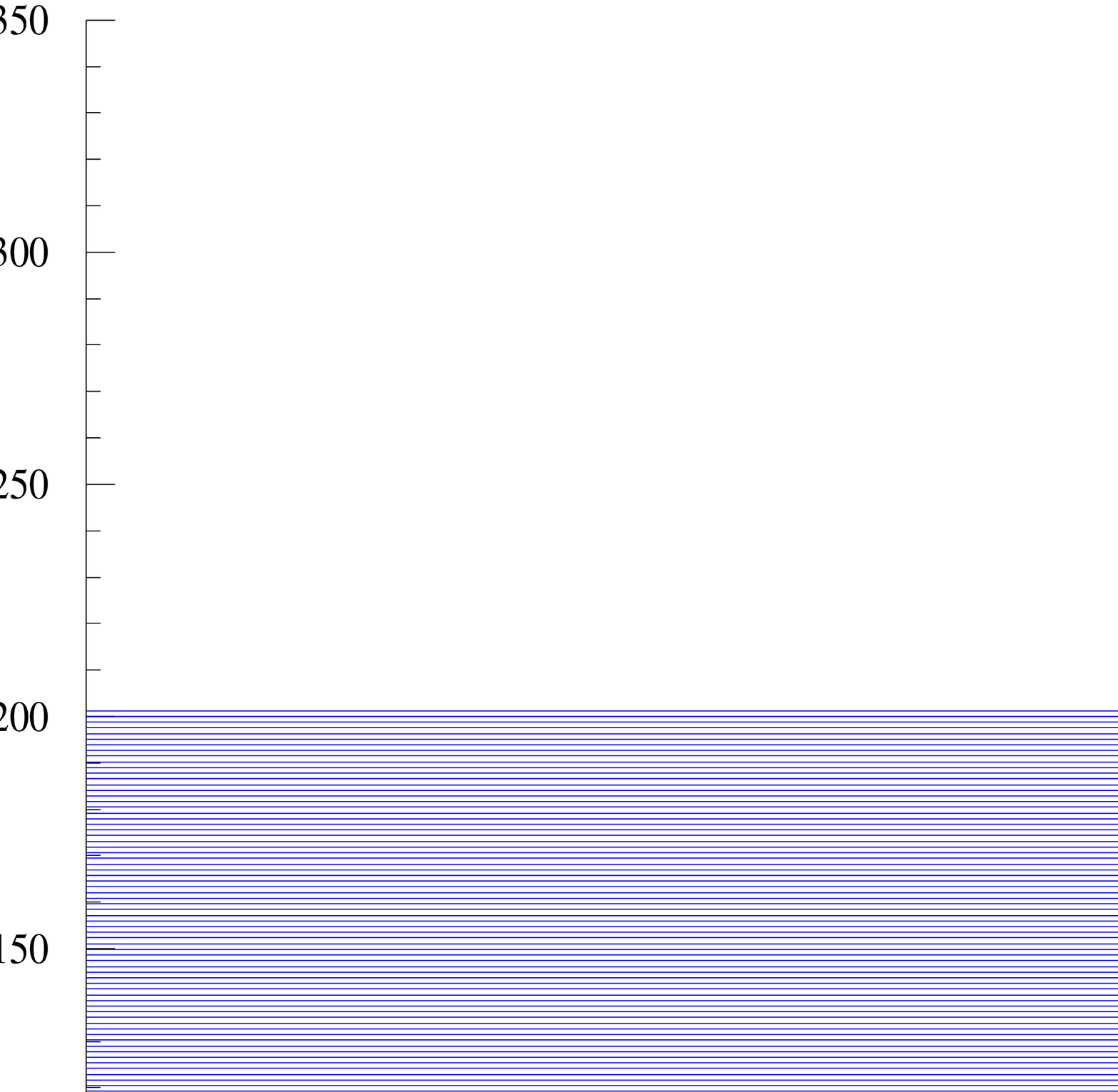,height=1.3in}
\psfig{figure=vtsum_f2b.ps,height=1.3in}
\caption{Tracking layout for the North American `L' (left) and
`S' (right) detector alternatives. Note the different
vertical and horizontal scales.
\label{fig:lands}}
\end{figure}

\section{Outline and Discussion of Proposed Tracking Alternatives}

An optimal tracking system has as large a radial extent as possible:
both inward, in order to optimize the measurement of the track origin
parameters (impact parameter), and outward, in order to provide as
large a lever arm as possible for the momentum measurement. A large
magnetic field is desirable, in order to provide maximum curvature
for the momentum measurement. Minimizing the single hit resolution,
both in absolute terms as well as normalized to radiation length,
improves the precision of the track reconstruction and limits the
effects of multiple scattering. It is interesting to note that the
central tracker plays an important role in the measurement of all five
tracking parameters. For example,
Figure~\ref{fig:cti} shows the $r-\phi$ impact parameter resolution
obtained~\cite{lcdtrk}
with and without the central tracker in the North American `S'
design (to be described below). It can be seen that
the angular measurement, and particularly the
curvature measurement, provided by the central tracker
are important for constraining the
track origin parameters.

While the European and Asian studies have focussed on a large volume
gaseous central tracker (TPC and drift chamber, respectively), the North
American community is considering two options, shown schematically
in Figure~\ref{fig:lands}: a large (`L') option with a TPC, and
a compact (`S') option with a central tracker composed of three
doublets of silicon drift or silicon $\mu$strip detectors. All
alternatives include a multi-layer silicon pixel inner tracker
for precise vertexing, with innermost layer at $r = 2.5$ cm ($1.2$ cm)
for the `L' (`S') alternative; ongoing studies indicate that
it may well be possible to reduce the `L' detector inner radius somewhat.
The three gaseous tracking alternatives are immersed in an axial
field of 2-3 T, while the `S' alternative assumes a 6 T field.
The `S' alternative includes a silicon
$\mu$strip forward tracking system; similar systems are currently
under study for the other three alternatives.
Intermediate tracking layers are under consideration
as well for the four alternatives.

\subsection{Tracking Performance}

\begin{figure}
\psfig{figure=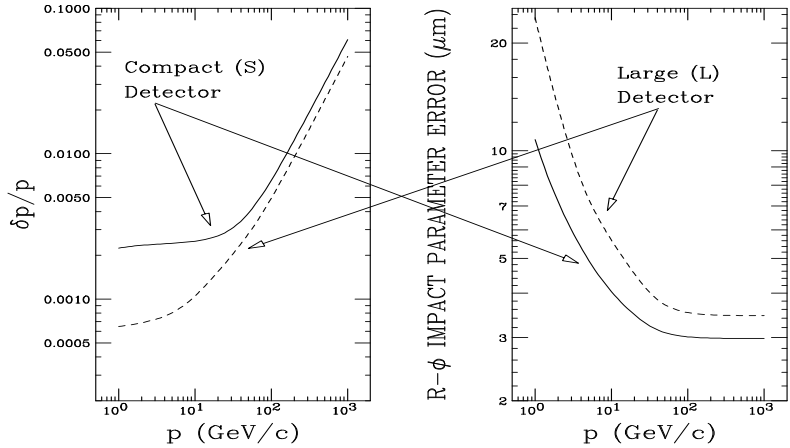,height=2.3in}
\caption{Comparison in momentum and impact parameter resolution for the
large and compact detectors.
\label{fig:dp}}
\end{figure}

Figure~\ref{fig:dp} compares the momentum resolution of the North
American `L' detector, fairly typical of the three gaseous tracking
alternatives, with that of the `S' detector. At high momentum, the
large magnetic field and excellent (10 $\mu$m) space-point accuracy
of the `S' detector compensates the longer lever arm of the `L'
detector, leading to a roughly equivalent momentum resolution.
At low momentum, the relatively low information content 
per radiation length of solid state tracking (1\% $X_0$ per layer
is assumed) leads to momentum resolution $\sim 3\times$ worse
for the `S' detector. Ideas relating to how this difference may
be lessened are sketched below; an important question to answer
via simulation studies is that of whether or not this difference in
momentum resolution has any impact on the physics to be studied.
Figure~\ref{fig:dp} also shows the $r-\phi$ impact parameter
resolution. The superior performance of the `S' detector is
primarily due to the proximity of the first measurement layer
to the interaction point.

\section{Vertexing Options}

The intense $e^+e^-$ pair background leads to an occupancy at
small radius as high as 1 hit/mm$^2$/bunch-crossing, leading
to the necessity of a pixel system for the innermost detector,
for which the fine granularity can permit the occupancy to be
low enough to do physics. One very attractive possibility is
CCD detectors. Current CCD vertex detectors, such the SLD
VXDIII detector~\cite{VXDIII},
feature a small ($20 \mu$m $\times$ $20 \mu$m) pixel
dimension, achieve better than $5 \mu$m resolution in both
$\phi$ and $z$, and are thin (0.4\% $X_0$ per layer).
On the other hand, though, the need to shift
the ionization signal by row and then column to the single
readout pad makes the CCD readout slow, and the CCD itself
susceptible to radiation damage. Active pixels, for which
the readout is bump bonded to the individual pixel sensors, are
both fast and radiation-hard, but are relatively thick, and do
not achieve as small a pixel dimension as do CCD's.
Time remains, however, to
address the shortcomings of these two approaches; the status
of ongoing pixel detector R\&D is presented below.

\subsection{CCD Detector Research and Development}

\begin{figure}
\psfig{figure=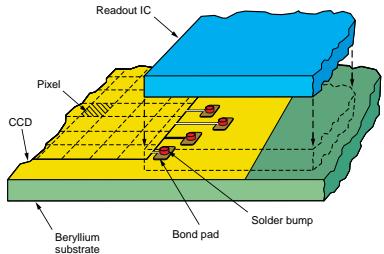,height=1.3in}
\caption{Schematic of the CCD column-parallel readout scheme.
\label{fig:cpr}}
\end{figure}

The two critical challenges to be met if CCD's are to
be used for inner tracking layers at the LC are the readout
speed and the radiation sensitivity. The SLD VXDIII has a
readout bandwidth of 25 MHz, which allows for the complete
pixel-by-pixel readout of the $8 \times 1.6$ cm CCD ladder
in approximately 200 msec. In the TESLA setting, with the
detector running in the proposed untriggered mode so that 
the beam crossing can not be inhibited during readout,
this leads to an integration over all $\sim 3000$ bunch crossings
in a single train, and thus an unacceptably high occupancy.
The LCFI collaboration in the UK~\cite{LCFI}
is currently exploring the possibility
a column parallel, rather than pixel-by-pixel, 
readout system (see Fig.~\ref{fig:cpr}).
This parallel readout, combined with a proposed readout
bandwidth of 50 MHz, would allow for the readout of the
LC CCD ladders in 25 $\mu$sec, integrating over only $\sim 75$
individual pulse crossings.

A group at the University of Oregon in the US has been studying the
mechanisms of radiation damage in hopes of devising a solution
suitable for the expected LC neutron fluence of approximately
$2 \times 10^{9}$/cm$^2$/yr in the inner layers of the
tracking system. Traps created by lattice deformations
caused by neutron scattering begin to reduce the efficiency
of the pixel-to-pixel charge transfer after approximately
$10^9$ neutrons/cm$^2$. The approach of the Oregon group
has been to fill these traps via a broad illumination of
the pixel array just prior to the arrival of the signal.
The diffusion of the electrons which thus fill the traps is
slowed by holding the array at cryogenic temperature
($\sim$190 K in the Oregon study). The Oregon group has
found that the neutron tolerance can be increased in this way
to greater than $10^{10}$ neutrons/cm$^2$, with some reason
to believe that the limit may be substantially higher.
In a LC CCD pixel tracker, this solution would be
implemented by clocking a single row of artificially introduced
charge through the pixel arrays just prior to the arrival of
beam at the interaction point.

In addition, the LCFI collaboration intends
to study the feasibility of further thinning the CCD substrate
while thickening the epitaxial (sensitive) layer. By making the
ladder thinner, it is hoped that the current material burden
of 0.4\% $X_0$ per layer can be reduced to as little as
0.12\% $X_0$ per layer. Thickening the epitaxial layer
would improve the overall S/N of the CCD array, allowing for
the possibility, with
appropriately optimized readout electronics, of single-hit
resolution of 3.5 $\mu$m or better.

\begin{figure}
\psfig{figure=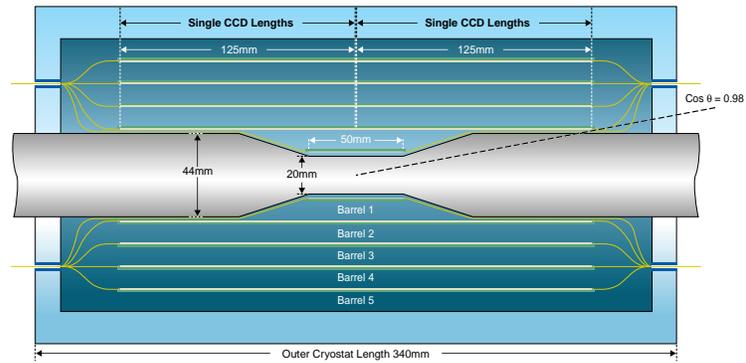,height=2.3in}
\caption{Proposed layout for the CCD Inner Tracker.
\label{fig:ccdet}}
\end{figure}

A possible layout of a CCD inner tracking detector is shown
in Fig.~\ref{fig:ccdet}, for the case of a 1 cm beampipe.
This layout provides 5-hit coverage
to $|\cos \theta| = 0.9$, and at least two hits out to
$|\cos \theta| = 0.96$.
The resolution of this device, for
$\cos \theta = 0$, has
already been exhibited in Fig.~\ref{fig:dp}. 
For the calculation
represented by this figure, a single-hit resolution of
5 $\mu$m was assumed, as well as a material burden of
0.12\% $X_0$ per layer.

\subsection{Active Pixel Sensor Research and Development}

\begin{figure}
\psfig{figure=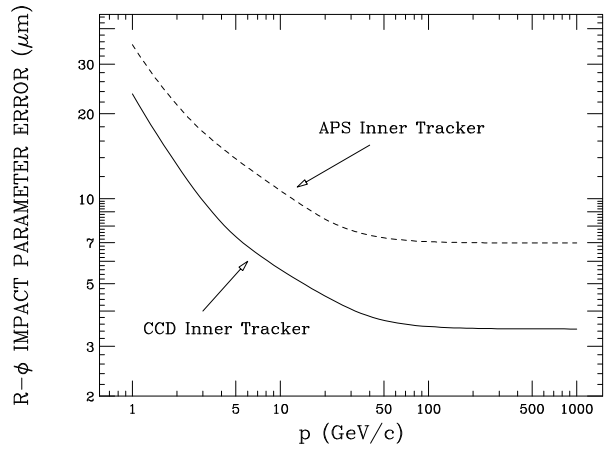,height=2.3in}
\caption{Comparison of the $r-\phi$ impact parameter resolution between
CCD and APS alternatives for the inner tracker.
\label{fig:ccdaps}}
\end{figure}

Active Pixel Sensor (APS) arrays, with individual readout electronics
bonded to each pixel site, avoid the need to transfer charge
pixel-by-pixel
to a common readout pad, and thus avoid the problems of radiation
hardness and readout speed encountered by CCD arrays. The development
of large-scale hybrid arrays, for which the pixel sensor and the
electronic readout are on separate chips which are connected
together by a multitude of precision bonds, is fairly advanced,
and is part of the baseline design of LHC detectors. However, existing
APS implementations
suffer from two disadvantages which detract from their desirability
for a LC detector -- the pixel dimension and the material
burden. The scale of the diffusion plume for the electron signal
in 300 $\mu$m of Si is $\sim 50 \mu$m. Thus, the pixel dimension
must be of this order or less to exploit charge sharing
in order to achieve a single-hit resolution of 10 $\mu$m or better.
On the other hand, the footprint of readout and driving circuitry
for the individual pixel channels has not yet been reduced to this
scale. In addition,
the hybrid approach, while proven, requires 
separate detector and readout planes,
as well as a high density of metallic bonds and readout and
cooling infrastructure.
For example, the ATLAS APS detector~\cite{ATPIX} has a pixel pitch
of $50 \times 300 \mu$m, and a material burden of $\sim 2\%$ $X_0$
per layer. With continued R\&D however, it has been suggested that
single-hit resolution as good as $10 \mu$m, with a material burden
of as little as $0.8 \%$ $X_0$ per layer, may be achieved for
an APS system.~\cite{ECFA}
A comparison of the $r-\phi$ impact parameter resolution
between the proposed CCD system and an APS system based on these
projections is shown in Fig.~\ref{fig:ccdaps}.

\begin{figure}
\psfig{figure=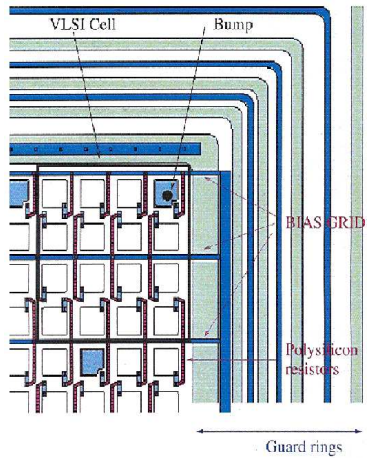,height=2.2in}
\caption{Schematic of the APS interleaved pixel layout.
\label{fig:interpix}}
\end{figure}

APS R\&D efforts are divided into two approaches, 
which are being explored by
a collaboration of groups from Milano, Cracow, Warsaw, Helsinki, and
Strasbourg, as well as a Hawaii-Stanford group in the US:
efforts to improve
upon existing hybrid implementations, and exploration of possible
monolithic approaches. The most promising hybrid approach, shown
in Fig.~\ref{fig:interpix}, involves interleaving pixels of dimension
$50 \mu$m $\times$ $50 \mu$m, but with a readout pitch of roughly
200 $\mu$m. Capacitive and resistive charge division would be
used to exploit the charge sharing between the uninstrumented pixels.
Prior studies with strip, rather that pixel, detectors with a 200 $\mu$m
readout pitch~\cite{intstrip}
indicate substantial resolution gain as the number of
interleaved strips is increased, with the single-hit resolution
improving from 77 $\mu$m with 0 interleaved strips to 9.7 $\mu$m with
3 interleaved strips. The European collaboration has received a set
of 10 wafers of interleaved pixel detectors, and has performed basic
I-V tests. They see no obvious flaws in the concept and layout, and
are in the process of designing a commensurate readout chip to allow
rigorous tests of the performance.

Monolithic APS detectors are intrinsically thinner devices, in that they
do not require a separate readout plane. In addition, not requiring
channel-by-channel bump bonding, they are mechanically simpler, and offer
the hope that the readout footprint can be sufficiently small to allow
direct charge-sharing measurements. Two prototype runs at Stanford's CIS facility
produced 1 mm$^2$ monolithic arrays with pixel dimensions of
$34 \mu{\rm m} \times 125 \mu{\rm m}$ (1993) and
$65 \mu{\rm m} \times 65 \mu{\rm m}$ (1996). Tests on the 1993 run
yielded a measured single-hit resolution of $2.0 \mu$m in the view for
which the pixel dimension was $34 \mu$m. Substantially larger arrays
($6 {\rm mm} \times 6 {\rm mm}$) are currently under development by
the European collaboration, although with a somewhat coarser (200-300 $\mu$m)
readout pitch, but have not yet been prototyped. Both the European and US
efforts are experiencing difficulty in interesting high-resistivity
commercial operations, necessary to obtain a substantial enough depletion
zone for minimum-ionization detection, in prototyping work. Questions have
also been raised regarding the scaling from small to large arrays,
for which intra-chip distances are large enough to require substantial
driving capabilities for the local readout electronics, thus potentially
restricting
the reduction of the readout pitch. Nevertheless, APS R\&D remains an
exciting avenue for development work, and is an active component of the
Linear Collider detector R\&D program.

\section{Options for the Central Tracker}

Challenges to be met in designing a central tracking system include
precise momentum resolution for the measurement of spectra endpoints
and invariant mass reconstruction, track separation capability in
dense Linear Collider jets, charge sign determination over a large range
in $|\cos \theta|$, the central tracker contribution to the impact
parameter resolution (see Fig.~\ref{fig:cti}), identification of
kinks in the decay of long-lived exotic states, dE/dX energy loss
for particle identification, and, not least, background immunity.
The relative weight that each of these different capabilities should
be awarded in design considerations is not yet clear, and is one of the
more important goals of the complementary physics simulation studies which
are now underway.

Four central tracking alternatives are
currently under consideration for the Linear Collider. These include
two gaseous tracking options (TPC and drift chamber), and two
solid-state tracking options (silicon $\mu$strip and silicon drift).
Discussed below are the some of the advantages and disadvantages
of each system, along with the associated R\&D programs, either
ongoing or proposed.

\subsection{The TPC Central Tracker}

Time Projection Chambers (TPC's) are a mature technology, and offer a
number of advantages for operation at a Linear Collider. The large
lever arm and minimal material budget (typically 1\% $X_0$ for the
gaseous detection medium; the TPC has no wires) that can be achieved
with a TPC provide superior momentum resolution over the full momentum
range of interest to the LC (c.f. Fig.~\ref{fig:dp}), even with a
moderate axial magnetic field of 2-3 T. The axial drift provides
excellent pixellation ($\sim 10^8$) for immunity to
background photon conversions, as well as track timing (from the offset
of the $z$-intercept) to within $\sim 5$ JLC/NLC beam crossings.
The numerous ($\sim 100$) measurement layers arrayed over a large
lever-arm provide excellent
redundancy for pattern recognition, a good dE/dX measurement, and
a substantial phase space for kink detection.

There are, however, a number of concerns associated with the use of
a TPC at a Linear Collider. The alignment tolerances required of
a 2m radius tracker making a momentum measurement of accuracy
$\delta p/ p^2 \; \simeq 5 \times 10^{-5}$ in a 2-3 T axial field
is of order $5 \mu$m. This specification applies to mechanical tolerances
as well as to space charge effects from drifting ionization due
to the gas multiplication at the readout planes and knock-ons from
background neutrons. Studies of shielding configurations as
well as the gas choice (light elements should be avoided) are
underway to mitigate the effects of the neutron background. Existing
TPC's make use of a conducting grid to inhibit the passage of gas-gain
ionization into the TPC drift region in between beam crossings; for
the short crossing interval of current LC designs, this gating
technique cannot be used. Instead, development work on position
sensitive detectors which intrinsically inhibit the release of
ions (discussed below in Section 7) are under development.

In addition, the track separation resolution for existing TPC's is
$\sim 2$ cm, although TPC experts expect that this can be reduced
to $\sim 5$ mm with some effort. This should be compared with
$\sim 100 \mu$m for solid-state trackers, although the need for
track separation resolution tends to decrease as the square of the
available lever-arm. Background photon ionization is not as well
localized in a gaseous tracker as for a solid state tracker, working
against background immunity (current estimates are for a 1\% occupancy
for the operation of a TPC in the LC environment). Finally, the end-cap
material burden may be an issue for forward instrumentation, which
is more critical for $e^+e^-$ collision at high energy due to the
enhanced effects of beamstrahlung and initial state radiation, as
well as the increased importance of peripheral and t-channel processes.

\subsection{The Drift Chamber Alternative}

The use of drift chambers as central tracking devices in colliding beam
physics is very well established. Nevertheless, several challenges
remain to establish the suitability of this alternative for the LC.
The Asian LC community has for now adopted the drift chamber
as its base-line alternative, and is looking into a number of issues,
primarily those having to do with the large radial and axial extent of
the proposed chamber. The chamber comprises 6 axial and 10 stereo
small jet-cell layers, with a radial extent from 40 to 230 cm, and
an axial length of 4.6 m. A prototype chamber has achieved resolutions
of 86 $\mu$m in $r-\phi$ and 1 mm in $r-z$, with a track separation
resolution of better than 2 mm. The group has paid particular attention
to gain variation caused by cell deformation due to the hyperboloid
wire surfaces introduced by the stereo angle, exacerbated by the
great length of the chamber. The resulting 0.43\% variation in the sense
wire surface field will result in a $< 10\%$ gas gain variation throughout
the length of the axial layers.
Thus, the Asian group has demonstrated
that problems related to gravitational sag and electrostatic deflection
can be adequately accounted and corrected for.

In a 2 T axial field, the chamber is expected
achieve a momentum resolution of $\sigma_p/p^2 =
1.1 \times 10^{-4}$, similar to that of the TPC alternative, and with
superior $r-\phi$ track separation performance. However, the lack
of axial segmentation generates some concern about the operation of
such a chamber in the LC environment. The operation of
the MARK-II and SLD
drift chambers at the SLC Linear Collider at SLAC, running at the
$Z^0$ pole, led to considerable restrictions on both the instantaneous
and integrated luminosity due to intolerable drift chamber occupancy.

\subsection{Silicon $\mu$strip Tracking at the LC}

The minute $r-\phi$ segmentation ($\sim 100 \mu{\rm m}$) and
excellent single-hit resolution (5-10 $\mu$m) of silicon
$\mu$strip tracking make it a natural candidate for operation in the
LC environment. The North American `S' detector (see Fig.~\ref{fig:lands})
includes a compact central tracker composed of 3 doublets of solid
state tracking (silicon $\mu$strip or silicon drift, the latter of which will be
discussed below) at radii between 14 and 71 cm, with an angular coverage
to $|\cos\theta| = 0.9$. As mentioned above, even with a 6 T axial field
the nominal momentum resolution
of such a system is not as good as that of the gaseous tracking alternatives,
suffering somewhat at high momentum from the lack of lever-arm and few
measurement layers, and substantially at
low momentum from the material burden (assumed
to be $1\%$ $X_0$ for each of the 6 measurement layers).
Plans for silicon $\mu$strip R\&D, to be described briefly below, lie
along two somewhat mutually exclusive paths: first, the design of
long ladders to minimize electronics and support structures, thus
reducing the material burden, and second, the speed-up of the $\mu$strip
readout to the allow timing to within a few NLC/JLC crossings, allowing a
factor of 10-50 improvement in the noise immunity of the tracker
in the NLC/JLC environment.

\begin{figure}
\psfig{figure=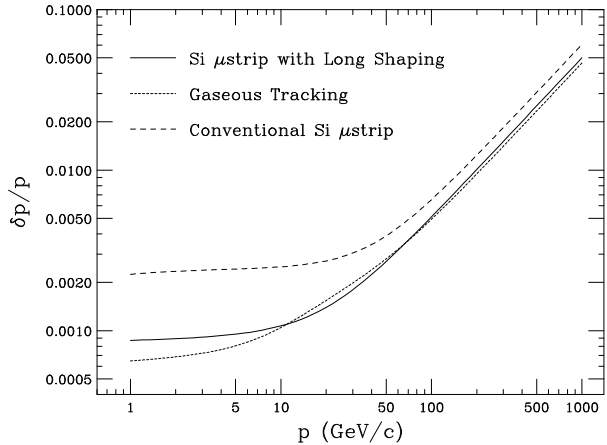,height=2.3in}
\caption{Comparison of expected momentum resolution between
long-shaping Si $\mu$strip, conventional Si $\mu$strip, and
gaseous tracking alternatives.
\label{fig:loscat}}
\end{figure}

Recent development of space-based silicon tracking systems has
put a premium on the low power consumption that can be achieved with
ultra-long $\mu$strip ladders. By exploiting a very long (2 $\mu$sec)
shaping time, the AMS collaboration~\cite{AMS} has developed $\mu$strip
preamplifier chip with a readout noise, in equivalent electrons, of
$$ \sigma_e \simeq 350 \; + \; 6 \; \cdot \; D $$
where $D$ is the ladder length in cm. Such a readout system would integrate
over all $\sim 100$ crossings of the NLC/JLC pulse train, and over less than
10 crossings of the TESLA pulse train; the estimated occupancy for a
50 $\mu$m pitch tracking system under these conditions is of order 0.1\%.
With a threshold of
$\sim 6000 e^-$, which yields $> 99\%$ efficiency for minimum-ionizing
particles traversing 300 $\mu$m of silicon at normal incidence,
ladder lengths approaching 200 cm
are conceivable -- long enough to instrument the full $|\cos \theta| < 0.9$
tracking acceptance at a radius of 70 cm
without having to introduce electronics and cooling
into the tracking volume. In addition,
detectors at smaller radius, where the ladder length and thus the readout
noise need not be so great, could be thinned. Finally, endcap-only support
structures, relying on the rigidity of the silicon itself for internal support,
may be possible,~\cite{HYTECH} allowing near-zero material burden for the
detector support. 

\begin{table}[t]
\caption{Layer Radii and Material Thickness for Long-Shaping Silicon
  $\mu$strip Tracker \label{tab:longs}}
\begin{center}
\begin{tabular}{|c|c|c|c|}
\hline
Layer \# &  Radius (cm)  & Axial Extent (cm) & Thickness (\% $X_0$) \\
\hline
\hline
1    &  14.0   & 29.0  & 0.11 \\
2    &  28.0   & 58.0  & 0.11 \\
3    &  42.0   & 87.0  & 0.21 \\
4    &  57.0   & 118.0  & 0.21 \\
5    &  71.0   & 147.0  & 0.32 \\
\hline
\end{tabular}
\end{center}
\end{table}

Table~\ref{tab:longs} shows the layer radii and material
thickness that might be achieved with this approach.
The resulting momentum resolution, shown in
Fig.~\ref{fig:loscat}, is essentially identical to that of gaseous tracker
systems. This system would not allow for the accurate timing of individual
tracks; however, accurate timing could be done by one or several layers
of fast gaseous tracking (to be described below) mounted at large radius, where
the associated material burden would not affect the track parameter
determination.

\subsection{Silicon Drift Detectors}

A variation on the concept of silicon $\mu$strip tracking is the use of
silicon drift detectors (SDD's). Although not as well established as $\mu$strip
technology, with the use of silicon drift detectors in the STAR vertex
detector at RHIC (approximately 1 m$^2$ of SDD's),
proponents of silicon drift detectors have demonstrated that
the technology is indeed plausible for a large-scale implementation.

\begin{figure}
\psfig{figure=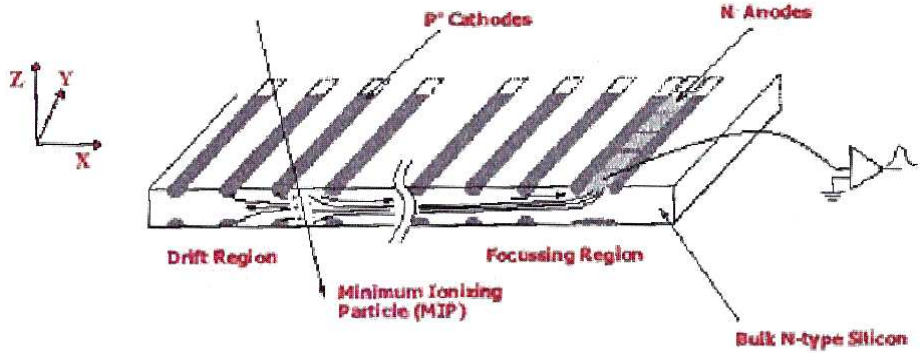,height=2.3in}
\caption{Schematic showing the principle of operation of the
Silicon Drift Detector.
\label{fig:sdd}}
\end{figure}

The principle of SDD's is shown schematically in Fig.~\ref{fig:sdd}.
Minimum-ionizing particles incident upon the detector produce
electron-hole pairs, just as for silicon $\mu$strip detectors.
However, in the case of SDD's, a pattern of azimuthal strips with
a graded potential drifts the electrons axially to a focusing
and collection region, which is segmented in azimuth to provide
$r-\phi$ information. Complementary $r-z$ information is available
from the drift time, giving the SDD a true 3-dimensional readout.
Single-hit resolutions achieved with the STAR system are better than
10~$\mu$m and 20 $\mu$m in $r-\phi$ and $r-z$
respectively, with track separation resolution
similar to that of silicon $\mu$strip devices.

Silicon drift detectors offer tremendous background immunity
for the LC environment. The pixel size, given for example by a 100 $\mu$m
strip pitch and 20 nsec digitization interval with a 6.5 $\mu$m/nsec drift,
would provide for $2.5 \times 10^{7}$ pixels in the
innermost layer of the LC detector. Tracks from minimum bias backgrounds
can be rejected, as for the TPC, by the axial drift provided they are
at least 3 JLC/NLC crossings (1 TESLA crossing) out of time.
SDD's are quite radiation hard, operating without degradation to a fluence
of approximately $10^{12}$ neutrons/cm$^2$, well in excess of the radiation
dose expected for the LC.

One possible disadvantage of SDD's, should ultra-precise momentum resolution
be deemed to be an important tracking system goal, is their incompatibility
with a long ladder implementation. The current material burden for the
STAR system is approximately 1.4\% per layer; with work, it may be possible
to reduce this to 0.5\% per layer.
It should be pointed out that SDD's have been operated in axial magnetic fields
as high as 6 T, the baseline value for the `S' detector alternative, with
the only degradation being a 10-15\% lower drift velocity due to
magneto-resistance. There are concerns, however, about the effect of
non-axial magnetic fields that arise as the boundary of the solenoid is
approached.

Proponents of SDD technology for the LC envision a program of developing
a 100 $\mu$m pitch detector (250 $\mu$m is now the standard) in hopes
of achieving a single-hit resolution of 5 $\mu$m. In order to reduce the
material burden, they would like to explore the possibility of
thinning the detectors from 300 to 150 $\mu$m, optimization of the hybrid
readout design, the use of gaseous rather than liquid cooling,
and the migration from a 4 inch to a 6 inch wafer for detector production.

\section{Specialized Gaseous Tracking Detectors}

Numerous groups in the particle physics community have been working
on the development of new types of gaseous tracking detectors,
including Gas Electron Multiplier (GEM) detectors, Micro-mesh
(Micromegas) detectors, and Micro-gap Wire Chambers (MGWC).
Several of these may well find application in a LC detector for
purposes of intermediate tracking, TPC readout, and timing layers.

\begin{figure}
\psfig{figure=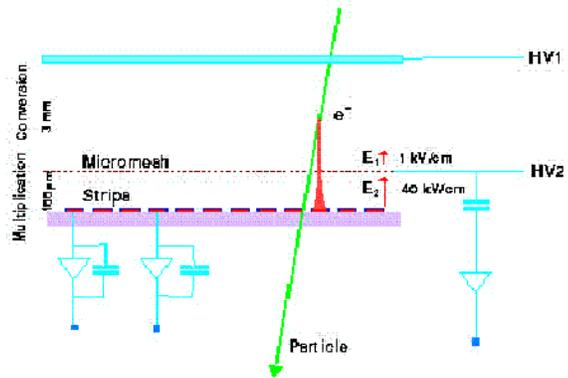,height=1.8in}
\caption{Schematic showing the principle of operation of the
Micromegas Detector.
\label{fig:mmgas}}
\end{figure}

The GEM detector consists of a matrix of holes in a thin ($\sim 50 \mu$m)
insulator sandwiched between metallic layers held at a potential
difference of several hundred volts. Gas gain is localized within
the region of the insulator, and residual ionization tends to be
captured by the low-potential metallic layer. It is also possible to
stack GEM detectors, leading to higher gains with greater ion feedback
suppression.

The Micromegas detector features a fine wire mesh separated by
$\sim 100 \mu$m (and several hundred volts) from an array of anode
strips on an insulated substrate (see Fig.~\ref{fig:mmgas}). A
several-millimeter thick conversion region, with a smaller
electrostatic field, provides a conversion layer for minimum-ionization
particles, and a drift towards the gas multiplication region beyond the mesh.
As for the GEM detector, the Micromegas detector suppresses the feedback
of liberated ions back into the conversion and drift region, in this
case via absorption by the mesh.

\begin{figure}
\psfig{figure=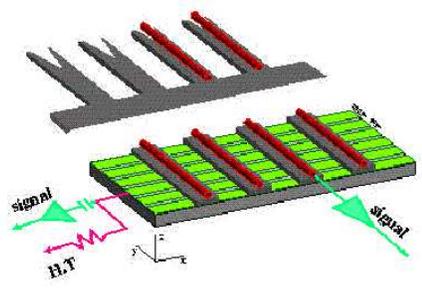,height=1.5in}
\caption{Schematic showing the principle of operation of the
Micro-gap Wire Chamber (MGWC).
\label{fig:mgwc}}
\end{figure}

Finally, the MGWC consists of an array of anode wires separated
from a plane of orthogonal cathode strips by a 10 $\mu$m strip
of insulator (see Fig.~\ref{fig:mgwc}).
Gas gain is achieved in the region surrounding the
anode wire. The readout is two-dimensional.

Table~\ref{tab:gasdet} exhibits some of the performance characteristics
of these three devices. As mentioned above, the GEM technology is a
candidate for the LC TPC readout, for which having a gating grid is
not feasible. With a 2 nsec temporal resolution, the MGWC might be a
candidate for a timing layer external to the central tracking system.
Finally,
the high rate capability of these specialized gas detectors might
make any one of them a candidate for the forward tracking system.
Work continues on the development of these detectors.

\begin{table}[t]
\caption{Performance characteristics of GEM, Micromegas, and MGWC
detectors. An asterisk ($\ast$) indicates that the parameter is
currently under study.  \label{tab:gasdet}}
\begin{center}
\begin{tabular}{|l|c|c|c|}
\hline
   &   GEM           & Micromegas            & MGWC                 \\
\hline
\hline
Gain                 &  $> 10^3$ per layer      & $> 10^4$               & $> 10^4$ \\
Spatial Resolution   &   45 $\mu$m              & 70 $\mu$m              & $\ast$   \\
Energy Resolution    &  $\pm 20 \%$             & $\pm 13 \%$            & $\ast$   \\
Timing Resolution    &   15 nsec                & 5 nsec                 & 2 nsec   \\
Rate Capability      & $> 10^5$ Hz/mm$^2$       & $> 10^5$ Hz/mm$^2$     & $\ast$   \\
Ion Feedback Supp.   &   $\sim 90 \%$ per layer & $\sim 90 \%$           & $\ast$   \\
\hline
\end{tabular}
\end{center}
\end{table}

\section{Forward Tracking}

With the large emphasis on t-channel and peripheral processes,
the forward tracking system is expected to play a critical role in
many LC physics analyses. Initial studies of the performance capabilities
of forward tracking have been done in the context of the North American
`S' detector, for which the forward tracking system is described
below, and for the European detector design.

It has been found that, even for far-forward angles approaching
$|\cos \theta| = 0.995$ (or 100 mrad, roughly the outer boundary
of the mask), the momentum resolution is dominated by the curvature
measurement (rather than the dip-angle measurement) over most of the
relevant momentum range. This has led to the selection of double-sided
silicon $\mu$strip disks for the `S' detector baseline, with opposing
small angle stereo ($\pm$ 20 mrad) on each side. A total of five disks
(five U and five V measurement layers) are spaced evenly from the
boundary of the CCD detector cryogenic containment to the endcap
calorimetry, for a total $z$ extent from 30 to 150 cm. A spatial
resolution of 7 $\mu$m per measurement, and a material burden of
1\% per layer, have been assumed. No concern has been given to the
LC backgrounds at very small radius; should this prove to be
problematic, it may be possible to use APS detectors for the
innermost annulus of each disk.

\begin{figure}
\psfig{figure=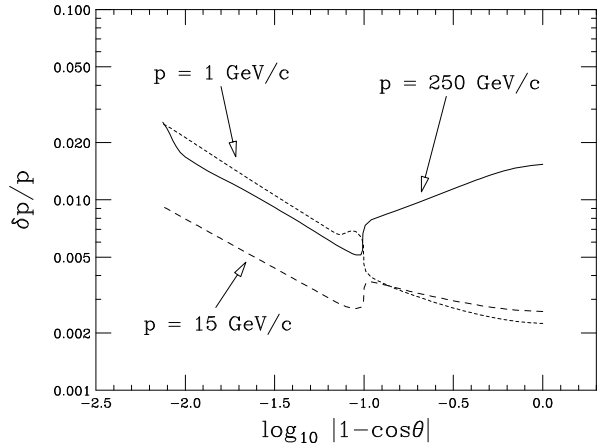,height=2.3in}
\caption{Expected momentum resolution for the North American `S'
 detector, for p = 1, 15, and 250 GeV/c, showing the performance
 in the forward region. A value of 
${\rm log}_{10} |1 - \cos \theta| = -2.3$ corresponds to
$\theta = 100$ mrad.
\label{fig:forward}}
\end{figure}

Figure~\ref{fig:forward} exhibits the fractional
momentum resolution for two
different momenta (1 and 250 GeV) for the `S' detector as a function
of $1 - |\cos \theta|$ for $\theta > 100$ mrad. It is seen that,
assuming perfect detector alignment, such a system can measure
the momentum to a relative precision of
better than 10\% over all values of momentum
and $|\cos \theta|$. In particular, the charge sign determination is robust
for even the smallest trackable angles.

\section{Summary and Conclusions}

Particularly for the central tracking detector, a number of
possible alternatives appear viable for a detector that would
instrument the interaction point of a high energy Linear Collider.
For the central tracker, the desire to have excellent momentum
and track separation resolution, as well as the highest possible
degree of background immunity, leaves no obvious choice for the
central tracker detector. It is essential that future physics studies
be directed towards determining what relative priority each of these
design criteria should have.

For the precise inner tracker, charge-coupled device (CCD) detectors
seem to have somewhat of an advantage in resolution and material
burden over active pixel sensor (APS) detectors. As discussed above,
though, several issues need to be resolved before CCD detectors can
be adopted to the exclusion of APS detectors. On the other hand,
APS detectors may well be the most satisfactory option for
far-forward, low radius tracking, and continued development of
APS detectors along the lines summarized herein should be encouraged.

Far-forward tracking has been presented in the context of the
North American `S' detector, and looks promising in terms of its ability
to perform a charge sign determination over the entire relevant range
of momentum and $|\cos \theta|$. More sophisticated studies, including the
effects of Linear Collider backgrounds, should be undertaken as early
as possible.

\section*{Acknowledgments}
The author would like to acknowledge the substantial 
help he received in writing this summary
from the contributors to the Tracking and Vertexing session
and from the organizers of the
Sitges workshop.
This work was supported in part by the Department of Energy,
Grant \#DE-FG03-92ER40689.



\section*{References}
\begin{thebibliography}{99}

\bibitem{daSilva} W. da Silva, these proceedings.

\bibitem{lcdtrk} Track parameter uncertainties have been
  calculated with the routine LCDTRK, available from SLAC at
  http://www-sldnt.slac.stanford.edu/nld/.

\bibitem{VXDIII} K. Abe {\it et al.}, Nucl. Instr. and Meth.
   {\bf A400}, 287 (1997).

\bibitem{ATPIX}  ATLAS Collaboration, Pixel Detector Technical Design Report,
http://atlasinfo.cern.ch/Atlas/GROUPS/INNER\_DETECTOR/PIXELS/tdr.html

\bibitem{ECFA}  S. Aid {\it et al.}, `Detector for the Linear Collider',
$e^+e^-$ Linear Collider Studies, Part E, DESY 97-123E, p. 509 (1997).

\bibitem{LCFI} LCFI Collaboration,
http://hep.ph.liv.ac.uk/~green/lcfi/lcfihome.html.

\bibitem{intstrip} M. Krammer, H. Pernegger, 
Nucl. Instr. and Meth. {\bf A397}, 232 (1997).

\bibitem{AMS} M. Pauluzzi (AMS Collaboration), Nucl. Instr. and Meth.
{\bf 383}, 35 (1996).

\bibitem{HYTECH} Bill Miller, HYTECH Inc., Los Alamos, New Mexico, USA,
Private Communication.

\end{thebibliography}
\end{document}